\documentclass{revtex4}

\usepackage{graphicx}
\usepackage{amsmath}
\usepackage{amssymb}
\usepackage{color}
\usepackage{epstopdf}

\begin{document}
\title{Thermophoresis of Brownian particles driven by coloured noise}
%\shorttitle{Thermophoresis of Brownian particles driven by coloured noise}

\author{Scott Hottovy}
\author{Jan Wehr}
\affiliation{University of Arizona, Tucson, Arizona 85721 USA}
\author{Giovanni Volpe}
\affiliation{Department of Physics, Bilkent University, Cankaya, Ankara 06800, Turkey}
%\shortauthor{S. Hottovy \etal}

%\pacs{05.60.-k}{Transport processes}
%\pacs{05.40.Jc}{Brownian motion}
%\pacs{47.61.-k}{Micro- and nano- scale flow phenomena}

\begin{abstract}
Brownian motion of microscopic particles is driven by collisions with surrounding fluid molecules. The resulting noise is not white, but coloured, due, e.g., to the presence of hydrodynamic memory. The noise characteristic time-scale is typically of the same order of magnitude as the inertial time-scale over which the particle's kinetic energy is lost due to friction.  We demonstrate theoretically that, in the presence of a temperature gradient, the interplay between these two characteristic time-scales can have measurable consequences on the particle's long-time behaviour. 
Using homogenization theory, we analyse the infinitesimal generator of the stochastic differential equation describing the system in the limit where the two time-scales are taken to zero keeping their ratio constant and derive the thermophoretic transport coefficient, which, we find, can vary in both magnitude and sign, as observed in experiments. Studying the long-term stationary particle distribution, we show that particles accumulate towards the colder (positive thermophoresis) or the hotter (negative thermophoresis) regions depending on their physical parameters.
\end{abstract}

\maketitle

\section{Introduction}\label{sec:intro}

A microscopic or nanoscopic object immersed in a fluid, e.g., a Brownian particle or a biomolecule, undergoes a permanent thermal motion. This motion is the result of the collisions with the fluid's molecules and is typically modelled as driven by a white Gaussian noise \cite{nelson}. However, this driving noise is actually coloured, i.e. it has a characteristic non-zero correlation time $\tau$, on a very short time-scale of the order of tens of nanoseconds, due, e.g., to the presence of hydrodynamic memory \cite{franosch2011}. This time-scale is similar to the particle's inertial relaxation time, i.e. the characteristic time for loss of kinetic energy through friction $\sigma = m/\gamma$, where $m$ is the mass of the particle and $\gamma$ the friction coefficient \cite{li2010}. As we will theoretically demonstrate, the interplay between these effects, despite occurring on time-scales that might not be in themselves directly accessible experimentally, can have measurable effects on the particle's long-time behaviour.

In this letter, we consider the dynamics of a Brownian particle driven by a coloured noise when it is immersed in a fluid where a temperature gradient is present. We find that due to the interplay between its two characteristic time-scales, $\tau$ and $\sigma$, the particle can exhibit a directed motion in response to the temperature gradient; furthermore, studying the long-term stationary particle distribution, we find that particles can accumulate towards the colder (positive thermophoresis) or the hotter (negative thermophoresis) regions depending on their physical parameters and, in particular, on the dependence of their mobility on the temperature. We remark that, as demonstrated by the mathematical analysis presented below, the presence of a coloured noise, as opposed to a white noise, is crucial for the emergence of such thermophoretic effects. The velocity of this motion can vary both in magnitude and sign, as observed in experiments \cite{platten2006,piazza2008b,srinivasan2011}. As a result, the particle moves either towards the cold or the hot side, quite similarly to what happens in the presence of an external driving force, e.g., gravity or electric fields; however, in this case, no \emph{external} force is actually acting on the particles \cite{piazza2008}.

\section{Mathematical model}\label{sec:model}

For a spherical particle of radius $R$ immersed in a fluid of viscosity $\mu$, which in general depends on the absolute temperature $T$, i.e. $\mu = \mu(T)$, the friction coefficient $\gamma$ satisfies Stokes law,
\begin{equation}\label{eq:gamma}
\gamma(T) = 6 \pi \mu(T) R,
\end{equation}
and the diffusion coefficient $D$ is related to $\gamma$ by the fluctuation-dissipation relation \cite{kubo},
\begin{equation}\label{eq:D}
D(T) = \frac{k_B T}{\gamma(T)}.
\end{equation}
For ease of argument, we will assume that the particle's motion is one-dimensional in a horizontal direction perpendicular to gravity, with position denoted by $x_t \in \mathbb{R}$ for all times $t\geq 0$.
We also assume viscosity to depend only on temperature, and fluid thermal expansion and convection to be negligible. If more complex models are needed, e.g. to account for interactions between the fluid and the particle or for the thermal expansion of the fluid, then the above model may need to be modified and the results may change, but the approach to address such problem will be the same as described in this Letter.
The relation in eq.~(\ref{eq:D}) assumes {\it local} thermodynamic equilibrium \cite{sengers}, which implicates that the temperature gradients should not be too steep; such conditions have been shown to be experimentally verified, e.g., in Ref.~\cite{duhr2006}.
The resulting motion of the particle is governed by the stochastic Newton equation:
\begin{equation}\label{eq:Newton}
m\ddot{x}_t = -\gamma(x_t)\dot{x}_t + \gamma(x_t) \sqrt{2 D(x_t)}\frac{\eta_t}{\sqrt{\tau}},
\end{equation} 
with initial conditions $x_0 = x^*$ and $\dot{x}_0 = v^*$.
The coloured noise $\eta_t$ is an Ornstein-Uhlenbeck process (OUP) defined by the stochastic differential equation (SDE):
\begin{equation}\label{eq:OUP}
d\eta_t = -\frac{2}{\tau}\eta_t\;dt + \sqrt{\frac{4}{\tau}}dW_t.
\end{equation}
where $\tau > 0$ is the noise correlation time and $W_t$ is a standard Wiener process. Its stationary solution is a zero-mean Gaussian process with $E[(\eta_t\eta_s)/\tau] = (1/\tau) \exp\left \{-\frac{2}{\tau}|t-s|\right \}$ and its covariance function converges to the delta function as $\tau$ tends to zero.

Typically, the hydrodynamic and inertial memory time-scales, i.e $\tau$ and $\sigma$, are very fast -- in particular, much faster than the typical time resolution at which the particle Brownian motion is experimentally sampled -- and their effects accumulate over the longer diffusive time-scale and eventually results in transport dynamics. Therefore, following Langevin's approach, it is customary to drop the inertial term, i.e. set the left hand side to zero in eq.~(\ref{eq:Newton}) \cite{nelson} and take the time correlation of the noise to zero. However, the limit of eq.~(\ref{eq:Newton}) as $m \rightarrow 0$, i.e. $\sigma \rightarrow 0$, has to be studied with care, requiring a nontrivial computation, as, in general, similar limits involve additional drift terms \cite{kupferman2004,freidlin2004,hottovy2012}. Here we are interested in the long-time behaviour of $x_t$ and in how the particle undergoes a deterministic drift in response to a temperature gradient as both the inertial and the noise characteristic times are taken to zero.
To this end, we will study eq.~(\ref{eq:Newton}) analysing the convergence of the infinitesimal operators of the corresponding diffusion processes (backward Kolmogorov equations), thus following the well-known methods from homogenization theory \cite{papanicolaou1975,schuss,pavliotis}. 
As $\sigma \rightarrow 0$ and $\tau \rightarrow 0$, we expect the limiting equation to be of the form:
\begin{equation}\label{eq:SKlimitSDE}
dx_t =A(x_t)\;dt + B(x_t)\;dW_t,
\end{equation}
where the effective coefficients $A(x)$ and $B(x)$ depend in general, as shown below, on the relative rate at which $\sigma$ and $\tau$ go to zero. Therefore, we define 
\begin{equation}\label{eq:theta}
\theta = \frac{\sigma}{\tau} = \frac{m}{\gamma \tau},
\end{equation}
which is a dimensionless parameter and, in general, depends on $x$, i.e. $\theta = \theta(x)$. We note that, for consistency, we choose to write the limiting eq.~(\ref{eq:SKlimitSDE}) with the It\^o convention -- a different choice would change the coefficients in the following discussion, but not the physical meaning. The effective drift term $A(x)$ reflects contributions resulting from the coloured nature of the noise, as well as from the small mass approximation. 
We stress that the particle's motion is perpendicular to gravity and $A(x)$ represents a real drift in the particle position, which arises despite the absence of external forces acting on it. This is a consequence of the system not being in thermodynamic equilibrium, as manifested by the presence of a temperature gradient \cite{piazza2008}.

For constant viscosity, i.e. $\mu(x) \equiv \mu_0$ and thus $\gamma(x) \equiv \gamma_0$, the simultaneous limit as $\tau \rightarrow 0$ and $\sigma \rightarrow 0$ was studied by Freidlin \cite{freidlin2004}: by taking $\tau \rightarrow 0$ first and then $\sigma \rightarrow 0$, he obtained the zero-mass limit eq.~(\ref{eq:SKlimitSDE}) with $A(x) =0$, while, by taking $\sigma \rightarrow 0$ first and then $\tau\rightarrow 0$, $A(x) = {B(x)B'(x) \over 2}$. Subsequently, Kupferman, Pavliotis and Stuart \cite{kupferman2004} showed that, if in the limit $\theta$ is constant in $x$, all the intermediate cases can be obtained as $\theta$ is varied. In a previous article\cite{hottovy2012}, we studied the limiting equation for a general class of equations similar to eq.~(\ref{eq:Newton}) but driven by a white noise. In this letter, we study the analogous problem for the coloured noise case.

\section{Multiscale analysis}\label{sec:multiscale}

To simplify the following analysis, we rewrite eq.~(\ref{eq:Newton}) as a set of first order SDEs, setting $m = \gamma \theta \tau$ [eq. (\ref{eq:theta})] and substituting $u_t^\tau = \sqrt{m}\dot{x}_t^\tau $:
\begin{equation}\label{eq:SDEgeneral}
\left\{
\begin{array}{rcl} 
dx_t^\tau &=& \frac{1}{\sqrt{\gamma \theta \tau}}u_t^\tau\;dt \\
du_t^\tau &=& \left [\frac{-1}{\theta \tau}u_t^\tau + \sqrt{\frac{2 D(x_t^\tau)\gamma(x_t^\tau)}{\theta}}\;\frac{\eta_t}{\tau}\right ]\;dt \\
d\eta_t &=& -\frac{2}{\tau}\eta_t \;dt + \sqrt{\frac{4}{\tau}}\;dW_t
\end{array}
\right.,
\end{equation}
with the initial conditions
$x_0^\tau = x^*$,
$u_0^\tau = v^*$
and $\eta_0$ a normal random variable with mean zero and variance ${1 \over  \tau}$ independent of the Wiener process $W_t$. The notation $x_t^\tau$ is introduced to distinguish the full 3-dimensional SDE with coloured noise and mass, from $x_t$ which, from now on, represents the limit as $\tau, \sigma \rightarrow 0$.

To determine the coefficients $A(x)$ and $B(x)$ for the limiting eq.~(\ref{eq:SKlimitSDE}), we use a multiscale analysis of the backward Kolmogorov equation associated with eqs.~(\ref{eq:SDEgeneral}) \cite{papanicolaou1975,schuss,pavliotis}: we will first write the backward Kolmogorov equation for eqs.~(\ref{eq:SDEgeneral}); we will then derive its limit as $\tau \to 0$ keeping $\theta$ constant; and, finally, we will determine $A(x)$ and $B(x)$ from this limiting Kolmogorov equation. 

Let $g(x',u',\eta',t'|x,u,\eta,t)$ be the probability density of the distribution of the position ($x'$), the rescaled velocity ($u'$) and the coloured noise ($\eta'$) of the particle at time $t'$ given their values $(x,u,\eta)$ at a time $t<t'$. The backward Kolmogorov equation for eqs.~(\ref{eq:SDEgeneral}) is:
\begin{equation}
\label{eq:Loperators}
\frac{\partial g}{\partial t} = \left[ \frac{1}{\tau}L_0 +\frac{1}{\sqrt{\tau}}L_1 \right] g.
\end{equation}
with 
$L_0 = \left[ \frac{-1}{\theta}u + \sqrt{\frac{ 2 D\gamma}{\theta}}\eta \right] \frac{\partial}{\partial u} - 2 \eta \frac{\partial}{\partial \eta} + 2\frac{\partial^2}{\partial \eta^2}$ and 
$L_1 = \frac{u}{\sqrt{\gamma\theta}}\frac{\partial}{\partial x}$.
We will take advantage of the fact that eq.~(\ref{eq:Loperators}) involves derivatives with respect to the $x, u,\eta, t$ variables only and shorten the notation by writing $g(x',u',\eta',t'|x,u,\eta,t) = g(x,u,\eta,t)$. Furthermore, we will not always explicitly indicate the dependence of $\gamma$, $D$ and $\theta$ on $x$.

Following the general multiscale analysis ansatz \cite{papanicolaou1975,schuss,pavliotis}, we postulate the solution to the Kolmogorov equation as a sum of an asymptotic series,
\begin{equation}
\label{eq:postulatedsolution}
g = g_0 + \sqrt{\tau} g_1 + \tau g_2 + ... ,
\end{equation}
and we substitute it in eq.~(\ref{eq:Loperators}). 
By matching corresponding powers of $\sqrt{\tau}$, we obtain the following equations:
\begin{align}
\label{eq:m}
L_0g_0 &= 0, \\
\label{eq:sqrtm}
L_0g_1 &= -L_1g_0, \\
\label{eq:unity}
\frac{\partial g_0}{\partial t} &= L_0g_2 + L_1g_1.
\end{align}

Eq.~(\ref{eq:m}) implies that $g_0(x,u,\eta) = g_0(x)$, since all derivatives of $L_0$ act on $u$ and $\eta$ only. 

A solution of eq.~(\ref{eq:sqrtm}) has the form, 
\begin{equation}
g_1(x,y,\eta) = \Phi(x,y,\eta)\frac{\partial g_0}{\partial x}, 
\end{equation}
where $\Phi$ solves the {\it cell problem} (see \cite{pavliotis} sec. 11.3) defined as 
\begin{equation}
\label{eq:cell}
-L_0 \Phi(x,u,\eta) = \frac{u}{\sqrt{\gamma\theta}}.
\end{equation}
We look for a solution of the form
\begin{equation}
\label{eqn:solut}
\Phi(x,u,\eta) = \left[ \sqrt{\frac{\theta}{\gamma}}u+\sqrt{2 D} \hat{h}(\eta) \right].
\end{equation}
Substituting eq.~(\ref{eqn:solut}) into eq.~(\ref{eq:cell}), we get the equation
$-2\eta \hat{h}_\eta +2\hat{h}_{\eta \eta} = \eta$ with a solution $\hat{h} = \eta/2$ and, therefore,
\begin{equation}
\Phi(x,u,\eta) = \sqrt{\frac{\theta}{\gamma}}u + \sqrt{\frac{ D}{2}}\eta.
\end{equation}

The solvability condition for eq.~(\ref{eq:unity}) follows from the Fredholm alternative \cite{pavliotis,courant,lax}:
\begin{equation}
\label{eq:solvability}
\int_{\mathbb{R}^2}\left \{-L_1 \left (\Phi(x,u,\eta)\frac{\partial g_0}{\partial x}\right ) + \frac{\partial g_0}{\partial t}\right \}\rho\;du\,d\eta = 0, 
\end{equation}
for all $\rho$ such that $L_0^*\rho=0$, 
where $L_0^*$ is the dual operator of $L_0$. 
Thus, we solve the dual problem
$L_0^*\rho = 0$
with the dual operator defined as
\begin{equation*}
L_0^* = -\frac{\partial }{\partial u}\left [ \left ( -\frac{1}{\theta}u + \sqrt{\frac{2 D\gamma}{\theta}}\eta \right )\cdot \right ]
+ 2 \frac{\partial }{\partial \eta}(\eta\cdot) +2 \frac{\partial^2}{\partial \eta^2}(\cdot).
\end{equation*}
This problem is equivalent to finding the stationary distribution for a two dimensional OUP in $u$ and $\eta$, with $x$ as a parameter, i.e. a zero-mean Gaussian probability density 
$\rho(u,\eta; x) = C(x)\exp\left \{ -\frac{1}{2}s_{11}u^2 -s_{12}u\eta -\frac{1}{2}s_{22}\eta^2\right \}$,
where $C(x)$ is a normalizing factor. Solving for the parameters $s_{ij}$ we obtain:
$s_{11} =\frac{1}{D\gamma} \left(\frac{1+2\theta}{2\theta}\right)^2$, 
$s_{12} = -\frac{1}{\sqrt{2 D \gamma \theta}} \frac{1+2\theta}{2\theta}$ and 
$s_{22} = \frac{1+2\theta}{2\theta}$.
Using this expression for $\rho$ we can evaluate the integral in eq.~(\ref{eq:solvability}) and obtain the backward Kolmogorov equation for $g_0$
\begin{equation}
\frac{\partial g_0}{\partial t} = A(x)\frac{\partial g_0}{\partial x} + B^2(x)\frac{\partial^2 g_0}{\partial x^2}, 
\end{equation}
where,
$A(x) = \int_{\mathbb{R}^2} \frac{\partial \Phi}{\partial x}\frac{u}{\sqrt{\gamma\theta}} \rho(u,\eta;x)\;du\,d\eta$
and 
$B(x)^2 = 2\int_{\mathbb{R}^2}\frac{u}{\sqrt{\gamma\theta}}\Phi\rho(u,\eta;x)\;du\,d\eta$.
Finally, we can compute the covariances with respect to the stationary density $\rho$, 
i.e. 
$E[u\eta] = \frac{\sqrt{2 D\gamma \theta}}{1+2\theta}$
and 
$E[u^2] = \frac{2\theta D \gamma}{1+2\theta}$,
and obtain explicit formulae for the coefficients of the effective SDE (\ref{eq:SKlimitSDE}):
\begin{equation}\label{eq:effectiveF}
A(x) =\frac{D'(x)\gamma(x)-4\theta(x) D(x)\gamma'(x)}{2\gamma(x)(1+2\theta(x))}
\end{equation}
and
\begin{equation}\label{eq:effectiveA}
B(x)^2 = 2 D(x).
\end{equation}
We remark that the diffusion term in the limiting eq.~(\ref{eq:SKlimitSDE}) is the same as in the initial eq.~(\ref{eq:Newton}).
The precise meaning of the limiting eq.~(\ref{eq:SKlimitSDE}) is that the solution $x_t^\tau$ of eq.~(\ref{eq:SDEgeneral}) converges in law to the solution $x_t$ of eq.~(\ref{eq:SKlimitSDE}) with the same initial condition $x_0 = x^*$ as $\tau,\sigma \rightarrow 0$. A rigorous proof is justified by techniques in \cite{Pardoux2003}.

\section{Physical interpretation: Drift and probability density}

Since the effective limiting SDE [eq.(\ref{eq:SKlimitSDE})] has been constructed using a stochastic integral with the It\^o convention, the expected position is
\begin{equation}
E[x_t] = E[x_0] + E\left [ \int_0^t A(x_s)\;ds\right ]. 
\end{equation}
The sign of $A(x)$ determines the direction towards which a Brownian particle is expected to travel. Therefore, if, e.g., we have $A(x)>0$, the particle will on average travel towards increasing $x$ until it reaches some boundaries, which can be either \emph{absorbing boundaries} or \emph{reflecting boundaries}.

\begin{figure}
\includegraphics[width=8.5cm]{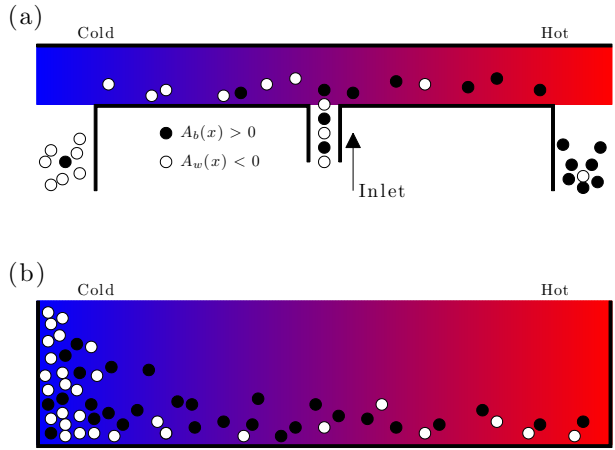}
\caption{(a) Schematic representation of the behaviour of thermophoretic particles in the presence of absorbing boundaries: 
black (white) particles have $A_b(x)>0$ ($A_w(x)<0$) [eq.~(\ref{eq:effectiveF})] and are mostly pushed towards the right (left) boundary, where they are removed from the channel.
(b) Schematic representation of the behaviour of thermophoretic particles in the presence of reflecting boundaries: the particles eventually reach a  steady state probability density $\rho_{\infty}(x)$ [eq.(\ref{eq:FPstationary})], which is different for particles with $A_b(x)>0$ (black) and $A_w(x)<0$ (white).}
\label{Fig1}
\end{figure}

In the presence of absorbing boundaries, particles disappear from the system as soon as they reach a boundary. Therefore, the sign of $A(x)$ determines the boundary at which particles are preferentially absorbed, as schematically shown in figure \ref{Fig1}(a).
This situation can be experimentally realized, for example, within a relatively long microfluidic channel where particles are steadily injected at a certain position, e.g., $x=0$, and removed once they reach either end of the channel. In the presence of a temperature gradient, these particles move towards increasing or decreasing $x$ depending on the sign of $A(x)$ and, therefore, can be sorted and classified on the basis of their physical and chemical properties that influence $A(x)$. Such a sorting process is  more efficient the longer the channel and the smaller the noise term, i.e. $B(x)$. 

In the presence of reflecting boundaries, particles are reflected back into the system when they reach a boundary. In this way, once a particle has interacted with the boundaries multiple times, the particle's position $x_t$ reaches a steady state probability density $\rho_{\infty}(x)$, as schematically indicated in figure \ref{Fig1}(b). The time-dependent $\rho(x,t)$ is the solution to the following Fokker-Planck equation (also known as forward Kolmogorov equation),
\begin{equation} \label{eq:FP}
\frac{\partial \rho}{\partial t} = -\frac{\partial}{\partial x}\left[ A(x)\rho\right] + \frac{\partial^2}{\partial x^2}\left[ \frac{B(x)^2}{2}\rho \right],
\end{equation}
with a given initial condition $\rho(x,0) = \rho_0(x)$. Since the Fokker-Planck equation is deterministic, its solution, i.e. the evolution of the probability density over time, does not involve any randomness. As $t \rightarrow +\infty$, the solution to eq.~(\ref{eq:FP}) converges to the steady state distribution $\rho_\infty(x)$, under certain conditions \cite[sec. 5.2]{risken}. 
If we assume the motion of the particle to be restricted to the interval $(a,b)$, $a<b$, then we can solve the stationary Fokker-Planck equation for the SDE~(\ref{eq:SKlimitSDE}) \cite{gardiner}. Given $A(x)$ and $B(x)$ the stationary solution is
\begin{equation}\label{eq:FPstationaryAB}
\rho_\infty(x) = \frac{C}{B(x)^2}\exp\left\{ 2 \int_a^x \frac{A(\tilde{x})}{B(\tilde{x})^2}\;d\tilde{x} \right\},
\end{equation}
which can also be expressed as a function of $D(x)$, $\gamma(x)$ and $\theta(x)$ as 
\begin{equation}\label{eq:FPstationary}
\rho_\infty(x) = CD(x)^{-\frac{1+4\theta(x)}{2+4\theta(x)}}\gamma(x)^{-\frac{2\theta(x)}{1+2\theta(x)}}, 
\end{equation}
where $C$ is a normalizing constant.
This situation can be experimentally realized in closed systems where a temperature gradient is present. Interestingly, this is the case of most experiments performed to study thermophoresis and the Soret effect, where a suspension of particles in a thermal gradient is given enough time to relax to its steady state distribution \cite{piazza2008,srinivasan2011}. 

\section{Results and discussion}\label{sec:results}

For the following discussion, we will express the thermophoretic drift as a function of $T$ and $\mu(T)$, {and since $\mu(T)$ is interpreted as an expansion around some temperature $T_0$, we will write $\Delta T = T(x)-T_0$, and $\frac{d}{dx}\mu(T) = \mu'(T)\Delta T$}. From eq.~(\ref{eq:effectiveF}), using eqs.~(\ref{eq:gamma}) and (\ref{eq:D}), we find that the effective thermophoretic drift is 
\begin{equation}\label{eq:thermodrift}
A(x) = k_B T'
\frac{\mu(T)-\mu'(T){\Delta T} \left[1+4\theta\right]}{12\pi R\mu^2(T) \left[1+2\theta\right]},
\end{equation}
where we have set $T = T(x)$ and $\theta = \theta(x)$, suppressing the dependence on $x$ for brevity and $\mu'(T)$ and $T'$ are derivatives with respect to $T$ and $x$ respectively. Eq.~(\ref{eq:thermodrift}) shows that the thermophoretic drift is determined not only by $T'$, but also by the dependence of $\mu$ on $T$. Interestingly, if $\mu'(T)>0$ and $\mu(T)>\mu'(T){\Delta T}$, from eq.~(\ref{eq:thermodrift}), there is a critical $\theta$ denoted $\theta_c$ such that the drift $A(x)$ changes sign:
\begin{equation}\label{eq:thetac}
\theta_c =  \frac{\mu(T)-\mu'(T){\Delta T}}{4\mu'(T){\Delta T}}.
\end{equation}

The stationary density is given as
\begin{equation}
\rho_\infty(x) = C\left [\frac{\mu(T(x))}{T(x)^{1+4\theta}}\right ]^{\frac{1}{2+4\theta}},
\end{equation}
which is an inverse power of $T$ unless $\mu(T)$ is at least quadratic in $T$.

Below, we consider in more detail the three simplest cases, i.e. $\mu$ constant, $\mu$ linear in $T$ and $\mu$ quadratic in $T$. We stress that, even though these are the first three orders of approximation, it might be necessary to consider more complex situations in real applications  because of the large range over which $T$ can vary.

\begin{figure}
\includegraphics[width=8.5cm]{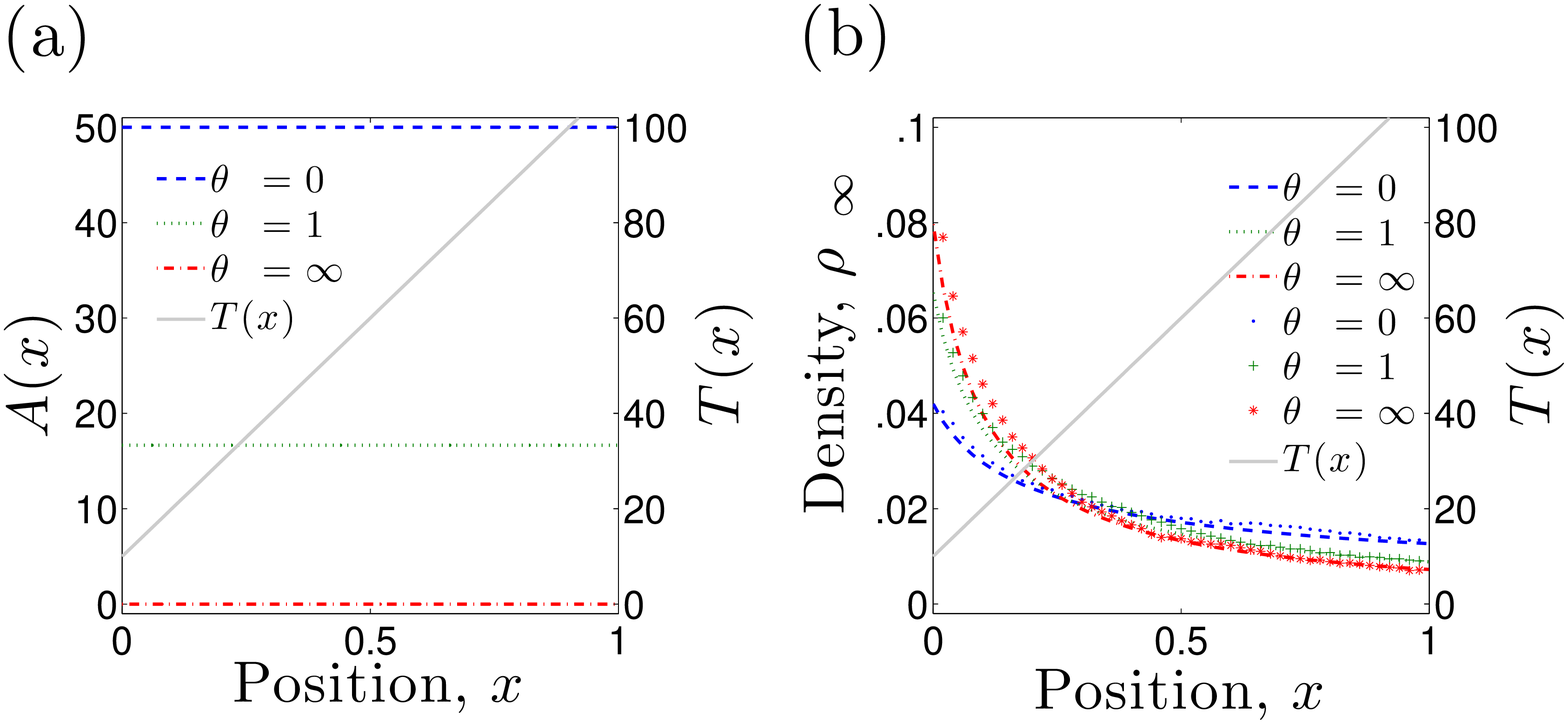}
\caption{$\mu(T)$ constant. (a) Thermophoretic drift $A(x)$ for various $\theta$ (dashed, dashed-dotted and dotted lines) in the presence of a temperature gradient (solid line). (b) Corresponding theoretical steady state distribution (lines) are always peaked towards the cold side. The symbols represent the steady states distributions resulting from Brownian dynamics simulations.}
\label{Fig2}
\end{figure}

\subsection{$\mu(T)$ constant}

In the simplest case $\mu$ does not depend on $T$, i.e. $\mu(T) \equiv \mu_0 > 0$, and the thermophoretic drift term (\ref{eq:thermodrift}) becomes
\begin{equation}\label{eq:thermodriftconst}
A(x) =\frac{k_BT'}{12\pi R\mu_0(1+2\theta)}.
\end{equation}
We stress that in this case $\theta$ is independent of $x$.
For example, let us consider the temperature gradient [grey solid line in figure \ref{Fig2}(a)]. The associated thermophoretic drift [eq.(\ref{eq:thermodriftconst})] is presented in figure \ref{Fig2}(a) for three values of $\theta$, corresponding to the situations where the time correlation of the coloured noise dominates ($\theta = 0$, blue dashed line), the time-scale of the particle inertia dominates ($\theta = +\infty$, red dotted--dashed line) and the two time-scales are comparable ($\theta = 1$, green dotted line). For all $\theta\geq 0$ the drift term has the same sign as $T'$, thus imposing a drift on the Brownian particle instantaneous flow towards the \emph{hotter} region. Equation~(\ref{eq:thermodrift}) features two terms, one term dependent on the frictional gradient and the second on the coloured noise; for $\mu$ constant, since there is no frictional gradient, the drift is only driven by coloured noise and the resulting flow increases with decreasing $\theta$.

If now the particles are allowed to interact repeatedly with the boundaries, they will eventually reach their stationary distribution, using eq.~(\ref{eq:FPstationary}), 
\begin{equation}\label{eq:statconst}
\rho_\infty(x)\propto T(x)^{ -\frac{1+4\theta}{2+4\theta} }.
\end{equation}
As shown in figure \ref{Fig2}(b), the particles will accumulate towards the areas of \emph{low} temperature for all $\theta\geq0$, which is in agreement with most experiments \cite{srinivasan2011}. This result is in striking contrast with the fact that the instantaneous thermophoretic drift actually pushes the particles in the opposite direction towards the \emph{hotter} regions. Such a difference between the instantaneous drift and the long-term stationary distribution has also been observed in systems at thermodynamic equilibrium \cite{lancon2001,Ao2007,volpe2010,brettschneider2011}.

\begin{figure}
\includegraphics[width=8.5cm]{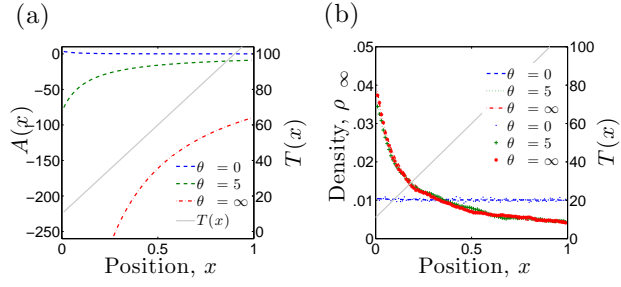}
\caption{Same as figure \ref{Fig2} for $\mu(T)$ linear. Note in (a) the sign change of $A(x)$ as $\theta$ crosses $\theta_c$.}
\label{Fig3}
\end{figure}

\subsection{$\mu(T)$ linear}

We now consider the case when $\mu(T)$ is linear, {i.e. $\mu(T(x))=\mu_0+\mu_1 \Delta T(x) > 0$}.  Interestingly, if $\mu_1>0$ (e.g. Ref. \cite{sharma2010}), $A(x)$ changes sign at a critical value of $\theta$ [eq.~(\ref{eq:thetac})]
\begin{equation}
\theta_c =\frac{\mu_0}{4\mu_1{\Delta T}}.
\end{equation}
The resulting effective thermophoretic drift (\ref{eq:thermodrift}) is shown in figure \ref{Fig3}(a) for the cases of $\theta = 0 <\theta_c$, $\theta = 1 >\theta_c$, and $\theta = \infty$. In particular,
\begin{equation}
\label{eq:theta0}
A(x) =\frac{k_BT'\mu_0}{12\pi R(\mu_0+\mu_1{\Delta T})^2} \mbox{ for }\theta=0
\end{equation}
and
\begin{equation}
\label{eq:thetainfty}
A(x) = -\frac{k_BT'\mu_1{\Delta T}}{6\pi R(\mu_0+\mu_1{ \Delta T})^2}  \mbox{ for }\theta \rightarrow +\infty,
\end{equation}
whose dependence on the thermal gradient clearly shows opposite signs if $\mu_1>0$.

For the stationary distribution [figure \ref{Fig3}(b)], 
\begin{equation}
\rho_\infty(x) \propto \left[ \mu_0{\Delta T(x)}^{-(1+4\theta)}+\mu_1{\Delta T(x)}^{-4\theta} \right]^{\frac{1}{2+4\theta}}. 
\end{equation}
For $\mu_1>0$ it is clear that $\rho_\infty$ will be an inverse power of $T(x)$. If $\mu_1<0$, since $\gamma(x)>0$ for all $x$, then $\mu_0>|\mu_1{\Delta T(x)}|$ and $\mu_0{\Delta T(x)}^{-(1+4\theta)}>|\mu_1{\Delta T(x)}^{-4\theta}|$. Thus $\rho_\infty$ will be an inverse power of temperature for all admissible $\mu_0,\mu_1$. This suggests that the particle will more likely be found in the \emph{colder} regions. Interestingly, unlike the expected drift, derived from eqs.~(\ref{eq:theta0})-(\ref{eq:thetainfty}), there is no qualitative change in behaviour at $\theta_c$.

\begin{figure}
\includegraphics[width=8.5cm]{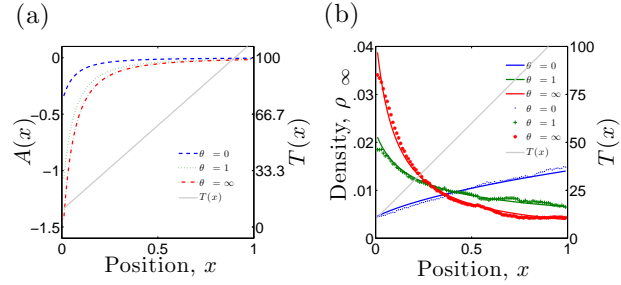}
\caption{Same as figure \ref{Fig2} for $\mu(T)$ quadratic. Note in (b) the change of the distribution peak from cold to hot as a function of $\theta$.}
\label{Fig4}
\end{figure}

\subsection{$\mu(T)$ quadratic}

We finally consider the case when $\mu(T)$ is quadratic in $T$, i.e. $\mu(T(x))=\mu_0+\mu_1 {\Delta T(x)}+\mu_2{\Delta T(x)}^2$.  Again, $A(x)$ changes sign at [eq.~(\ref{eq:thetac})]
\begin{equation}
\theta_c =\frac{\mu_0-\mu_2{\Delta T}^2}{4(\mu_1{\Delta T}+2\mu_2{\Delta T}^2)},
\end{equation}
if $\mu_0>\mu_2{\Delta T}^2$ and $\mu_1{\Delta T}+2\mu_2{\Delta T}^2>0$, or $\mu_0<\mu_2{\Delta T}^2$ and $\mu_1{\Delta T}+2\mu_2{\Delta T}^2<0$.
The resulting effective thermophoretic drift (\ref{eq:thermodrift}) for the extreme cases is
\begin{equation}
A(x) =\frac{k_BT'(\mu_0 - \mu_2{\Delta T}^2)}{12\pi R(\mu_0+\mu_1{\Delta T}+\mu_2{\Delta T}^2)^2} \mbox{ for }\theta=0
\end{equation}
and
\begin{equation}
A(x) = -\frac{k_BT'(\mu_1{\Delta T}+2\mu_2{\Delta T}^2)}{6\pi R(\mu_0+\mu_1{\Delta T}+\mu_2{\Delta T}^2)^2}  \mbox{ for }\theta \rightarrow +\infty,
\end{equation}
whose dependence on the thermal gradient shows opposite signs for $\mu_0>\mu_2{\Delta T}^2$. In figure \ref{Fig4}(a), we study the case $\mu_0<\mu_2{\Delta T}^2$ and $\mu_1{\Delta T}+2\mu_2{\Delta T}^2$.

The stationary distribution is
\begin{equation}
\label{eq:statquad}
\rho_\infty \propto \left[ \mu_0{\Delta T}^{-(1+4\theta)}+\mu_1{\Delta T}^{-4\theta} + \mu_2{\Delta T}^{(1-4\theta)} \right]^{\frac{1}{2+4\theta}},
\end{equation}
where it is clear that there are combinations of $\mu_0$, $\mu_1$ and $\mu_2$ for which the density will incur a transition from peaking at colder regions to hotter, as shown in figure \ref{Fig4}(b). This may lead to interesting behaviours as a function of the temperature. For example, in reference \cite{duhr2006pnas}, DNA particles ($\theta \ll 1$) change from accumulating in colder regions when the minimum temperature is {$T = 276$ K to accumulating in warmer regions when a new minimum temperature $T = 293$ K and the gradient is left unchanged ($2$ K between the colder and warmer regions). 
In agreement with this experiment we predict, for $T = 276$ K, since, {expanding $\mu(T)$ around $T_0 = 273$}, $\mu_0 + \mu_1{\Delta T}> \mu_2{\Delta T}^2$, the stationary distribution $\rho_\infty$ [eq. (\ref{eq:statquad})] peaks in colder regions for all $\theta$, and, for $T=293$ K, since $\mu_2 {\Delta T}^2>\mu_0 + \mu_1{\Delta T}$, the stationary distribution peaks in hotter regions for $\theta<1/4$, which is verified in this case.}{ However, we note that the model in this paper leads to a different explanation than the one in reference \cite{duhr2006pnas} and there may be numerous other factors influencing thermophoresis including thermal expansion of the fluid and convection.}

\section{Conclusions and outlook}\label{sec:conclusions}

In this letter, we have given a systematic analysis of a system with two short time-scales: the time correlation of the coloured noise $\tau$ and the inertial relaxation time $\sigma$. We have derived the effective thermophoretic drift $A(x)$ and the steady-state probability distribution $\rho_{\infty}(x)$ in the limit as these time-scales go to zero. We have finally applied these results to study the thermophoretic motion of a  Brownian particle in a temperature gradient, showing how, in agreement with experiments, $\rho_{\infty}(x)$ tends to be peaked towards the colder region (positive thermophoresis), but can switch to hotter regions (negative thermophoresis) under the right conditions. As possible future lines of research, also noises that are not OUP can be considered: these might be particularly promising to model the effects of the chemical interaction between a particle and a solvent, of the viscoelasticity of the medium or of depletion forces due, e.g., to polymers or small particles in the solution.

\acknowledgments
The authors thank Austin McDaniel for his critical reading of the manuscript. SH was supported by the VIGRE grant through the University of Arizona Applied Mathematics Program. GV was partially supported by TUBITAK grant 111T758. JW and SH were partially supported by the NSF grant DMS 1009508.

%%\ibliographystyle{eplbib}      % mathematics and physical sciences
\bibliographystyle{spphys}       % APS-like style for physics
\bibliography{sde_bib2}   % name your BibTeX data base

\end{document}